\documentclass[prl, twocolumn, superscriptaddress,floatfix, nofootinbib]{revtex4-1}

\usepackage{url}
\usepackage{xspace}

\usepackage{amssymb}
\usepackage{amsmath}
\usepackage{graphicx}
\usepackage[small]{subfigure}

\usepackage{color}
\definecolor{darkred}{rgb}{1.0,0.1,0.1}
\definecolor{darkgreen}{rgb}{0.1,0.7,0.1}
\definecolor{darkblue}{rgb}{0.1,0.1,1.0}

\DeclareRobustCommand{\Fig}[1]{Fig.~\ref{fig:#1}}

\DeclareRobustCommand{\Eq}[1]{Eq.~(\ref{eq:#1})}

\newcommand{\eqn}[1]{\begin{align}#1\end{align}}

\def\cL{\mathcal{L}}
\def\cJ{\mathcal{J}}

\newcommand{\Pythia}{{\sc Pythia}\xspace}
\newcommand{\Herwig}{{\sc Herwig}\xspace}

\newcommand{\JUNIPR}{{\sc Junipr}\xspace}
\newcommand{\JUNIPRtitle}{JUNIPR\xspace}

\begin{document}

\title{Binary \JUNIPRtitle: an interpretable probabilistic model for discrimination}

\author{Anders Andreassen}
\email{andersja@berkeley.edu}
\affiliation{Department of Physics, University of California, Berkeley, CA 94720 \\ and
Theoretical Physics Group, Lawrence Berkeley
National Laboratory, Berkeley, CA 94720}

\author{Ilya Feige}
\email{ilya@faculty.ai}
\affiliation{Faculty, 54 Welbeck Street, London, W1G 9XS}

\author{Christopher Frye}
\email{chris.f@faculty.ai}
\affiliation{Faculty, 54 Welbeck Street, London, W1G 9XS}

\author{Matthew D.~Schwartz}
\email{schwartz@physics.harvard.edu}
\affiliation{Department of Physics, Harvard University, Cambridge, MA 02138}

\begin{abstract}
\JUNIPR is an approach to unsupervised learning in particle physics that scaffolds a probabilistic model for jets around their representation as binary trees.  
Separate \JUNIPR models can be learned for different event or jet types,
then compared and explored for physical insight.
The relative probabilities can also be used for discrimination. 
In this paper, we show how the training of the separate models can be refined in the context of classification to optimize discrimination power. 
We refer to this refined approach as Binary \JUNIPR. 
Binary \JUNIPR achieves state-of-the-art performance for quark/gluon discrimination and top-tagging. 
The trained models can then be analyzed to provide physical insight into how the classification is achieved. 
As examples, we explore differences between quark and gluon jets and between gluon jets generated with two different simulations.
\end{abstract}

\maketitle

Modern machine learning has already made impressive contributions to particle physics. Convolutional~\cite{deOliveira:2015xxd, Baldi:2016fql,Komiske:2016rsd,Komiske:2017ubm,Kasieczka:2017nvn, Macaluso:2018tck}, recurrent and recursive networks~\cite{Butter:2017cot,Louppe:2017ipp,Cheng:2017rdo,Egan:2017ojy, Fraser:2018ieu}, autoencoders~\cite{Collins:2018epr,Farina:2018fyg,Blance:2019ibf,Roy:2019jae}, adversarial networks~\cite{Paganini:2017dwg,deOliveira:2017pjk,Paganini:2017hrr} and more have been shown effective in applications including
quark/gluon jet discrimination, top-tagging and pileup removal.  A key question that is beginning to be addressed is: what is the optimal
representation of the information in an event? Is it through analogy with images~\cite{deOliveira:2015xxd,Baldi:2016fql}, natural-language processing~\cite{Louppe:2017ipp,Fraser:2018ieu}, or set theory~\cite{Komiske:2018cqr, Qu:2019gqs}? In many of these approaches, there is a competition between effectiveness in some task (e.g.~pileup removal, jet classification) and interpretability of the neural network. An approach to machine learning for particle physics called \JUNIPR \cite{JUNIPR} builds a separate network for each jet type using a physical representation of the information in the jet: the jet clustering tree. In~\cite{JUNIPR} a method for  construction and training of such a network was introduced. In this paper, we show how the \JUNIPR framework can be used in discrimination tasks, achieving state-of-the art classification power while maintaining physical interpretability.

\begin{figure}[t]
\centering
\includegraphics[width=\linewidth]{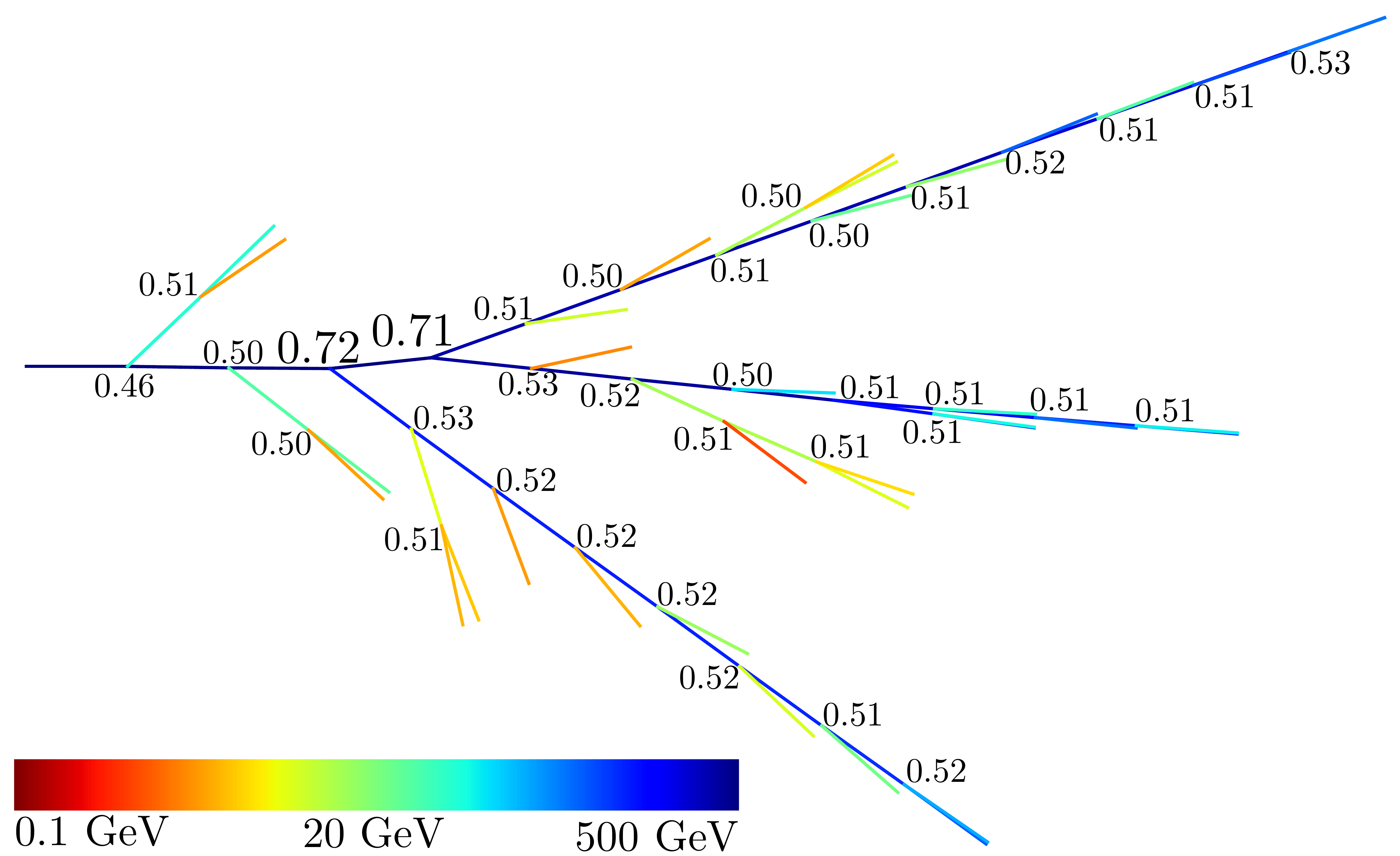}
\caption{Binary \JUNIPR tree for jet drawn from mixed top/QCD test set. Binary \JUNIPR predicts ``top'' with high probability: $P_\cJ(\text{top}\,|\,\text{jet}) = 0.99$. 
Each node is labeled with the probability that the jet is top-type, given only the information at that branching.
Planar angles correspond to 3D opening angles between clustered momenta, and color corresponds to energy.
The final factor corresponding to the tree's true end is not shown: $P^{(n)} = 0.52$; see \Eq{Pend}.
}
\label{fig:jet_trees}
\end{figure}

\JUNIPR begins by taking each jet in some sample and clustering it into a binary tree according to some deterministic algorithm. 
See \Fig{jet_trees} below for an example of such a tree.
The algorithm can be physically motivated (like the $k_T$~\cite{Ellis:1993tq} or Cambridge/Aachen~\cite{Dokshitzer:1997in} algorithms) but does not have to be. In such a tree, the momenta of each mother branch is the sum of the momenta of her daughters.
We denote the momenta
of the particles in the jet by $\{p_1\ldots p_n\}$ and the momenta in the clustering tree by $\{k_1^{(t)}\!\ldots\, k_t^{(t)}\}$ at branching step $t$. To be concrete, at $t=1$ we have $k_1^{(1)} = p_1 + \cdots + p_n$, at $t=n$ we have $\{k_1^{(n)} \!\ldots\, k_n^{(n)}\} = \{p_1 \ldots p_n\}$, and at each branching in between, $\{k_1^{(t)} \!\ldots\, k_t^{(t)}\} \to \{k_1^{(t+1)} \!\ldots\, k_{t+1}^{(t+1)}\}$  involves a single $1 \to 2$ momentum splitting. \JUNIPR learns to compute the probability $P_\cJ$(jet) of the jet, meaning the probability that the corresponding set of final state momenta $\{p_1\ldots p_n\}$ would be found in the given sample.
This probability can be factorized as a product over branching steps in the clustering tree:
\eqn{
\nonumber
P_\cJ(\text{jet}) &= \Bigg[\prod_{t=1}^{n-1} 
P^{(t)}\big(k^{(t+1)}_1\!\ldots\, k^{(t+1)}_{t+1} \big| k^{(t)}_1\!\ldots\, k^{(t)}_{t}\big)
\Bigg] 
\\&\times 
P^{(n)}\big(\text{end} \big| k_1^{(n)} \!\ldots\, k_n^{(n)}\big)
\label{eq:BasicProductAssumption}
}

To learn these probability distributions, \JUNIPR introduces a quantity $h^{(t)}$ as a representation of $\{k^{(t)}_1\!\ldots\, k^{(t)}_{t}\}$, i.e.~the ``state'' of the jet at branching step $t$. 
\JUNIPR learns to compute $h^{(t)}$ in training.  In machine-learning language, $h^{(t)}$ is the autoregressive latent variable, which in our implementations is taken to be the latent state of a recurrent neural network. Then we can write, e.g.,
\begin{equation}
    P^{(n)}\big(\text{end} \big| k_1^{(n)} \ldots k_n^{(n)}\big) = 
P_\text{end}\big(\text{true}\big|h^{(n)}\big) \label{eq:Pend}
\end{equation}
where $P_\text{end}(\text{true}|h^{(n)})$ is the binary probability that the clustering tree ends at branching step $n$.

\JUNIPR further factorizes the branching-probabilities of \Eq{BasicProductAssumption} into more intuitive probability distributions
\eqn{
\label{eq:IndividTimeStepModel}
P^{(t)}\big(k^{(t+1)}_1\!\ldots\big| k^{(t)}_1\!\ldots\big)
&=
P_\text{end}\big(\text{false}\big|h^{(t)}\big)\\
&\times
P_\text{mother}\big(m^{(t)}\big|h^{(t)}\big)\nonumber \\
&\times
P_\text{branch}\big(k_{d_1}^{(t+1)} k_{d_2}^{(t+1)} \big| k_m^{(t)} \, h^{(t)}\big) \nonumber
}
Here, $P_\text{end}(\text{false}|h^{(t)})$ is the binary probability that the clustering tree does not end at branching step $t$, $P_\text{mother}(m^{(t)}|h^{(t)})$ is the discrete probability that tree-momentum $k_m^{(t)}$ will participate in the $1 \to 2$ branching at step $t$, and $P_\text{branch}\big(k_{d_1}^{(t+1)} k_{d_2}^{(t+1)} \big| k_m^{(t)}\, h^{(t)}\big)$ is the distribution over daughters of the branching. 
Structuring the probabilistic model in terms of the product of these parts is essential to 
the interpretability of the model's output, as each part has separate physical meaning.

The latent state $h^{(t)}$ has access to the global content of the jet at branching step $t$, i.e.~to all the momenta $\{k_1^{(t)}\!\ldots\, k_t^{(t)}\}$. The factorization over branching steps is powerful, and useful to the extent that the $1 \to 2$ branching dynamics encoded in $P_\text{branch}\big(k_{d_1}^{(t+1)} k_{d_2}^{(t+1)} \big| k_m^{(t)} \, h^{(t)}\big)$ are local,
depending only weakly on $h^{(t)}$.
Even if there were no evidence for this factorization in the training data (as was explored with ``printer jets'' in~\cite{JUNIPR}), \JUNIPR would still learn the probability distributions, but physical interpretability would be lost. 

In \cite{JUNIPR}, \JUNIPR was trained to model jet dynamics via unsupervised learning. In that approach, the probabilistic model is learned by maximising the log likelihood of $P_\cJ$ over the training data:
\eqn{
\text{log likelihood} = 
\sum_{\text{jets}}
\; \log
P_\cJ \big( \text{jet} \big)
\label{eq:UnsuperObjective}
}
where the sum is over jets $\{p_1\ldots p_{n}\}$  in the training set. We call this the ``unary objective function''.
Despite being unsupervised, this approach can be used to discriminate between two jet types, say $a$ and $b$. To accomplish this, one trains two separate \JUNIPR models: $P_\cJ(\text{jet}|a)$ on a data set containing predominantly type-$a$ jets and $P_\cJ(\text{jet}|b)$ on predominantly type-$b$ jets. Discrimination between $a$ and $b$ is then achieved by thresholding the likelihood ratio $P_\cJ(\text{jet}|a) / P_\cJ(\text{jet}|b)$.

While discrimination by likelihood ratio is theoretically optimal in the perfect-model limit,
it has been shown that deep neural networks classify out-of-distribution data poorly \cite{nguyen2015deep, bendale2016towards}. 
That is, e.g., the $P_\cJ(\text{jet}|a)$ model is not expected to behave well on type-$b$ jets.
It is thus advantageous in practice to refine the training for discrimination.
By training directly for discrimination,  \JUNIPR can also focus model capacity on learning the often-subtle differences between type-$a$ and type-$b$ jets.
In fact, \JUNIPR's probabilistic nature makes supervised discrimination-learning very straightforward.
Assuming a mixed sample of both jet types, the probability that a given jet drawn at random belongs to class $a$ is, through Bayes' theorem, given by
\eqn{
    P(a|\text{jet}) = \frac{P(\text{jet}|a) \, P(a)}{P(\text{jet})}
}
For binary discrimination, $P(a|\text{jet}) + P(b|\text{jet}) = 1$, so 
\eqn{
    P(a|\text{jet}) = \frac{P(\text{jet}|a) \, P(a)}{P(\text{jet}|a) \, P(a) + P(\text{jet}|b) \, P(b)}
\label{eq:binary_pred}
}
Here $P(a)$ and $P(b)$ are simply the composition fractions $f_a$ and $f_b$ of the mixed sample, while $P(\text{jet}|a)$ and $P(\text{jet}|b)$ can be computed using two separate \JUNIPR networks as laid out in the paragraphs above.
This leads directly to the binary cross-entropy objective function one should use to train \JUNIPR for discrimination:
\eqn{
\cL 
&= \sum_{\text{a-jets}} \;
\log \frac{P_\cJ(\text{jet}|a) \, f_a}{P_\cJ(\text{jet}|a) \, f_a + P_\cJ(\text{jet}|b) \, f_b} \nonumber\\
&+ \sum_{\text{b-jets}} \;
\log \frac{P_\cJ(\text{jet}|b) \, f_b}{P_\cJ(\text{jet}|a) \, f_a + P_\cJ(\text{jet}|b) \, f_b}
\label{eq:SuperObjective}
}
where the sums extend over type-$a$ and type-$b$ jets in the training data, respectively. We call training with this objective function ``Binary {\JUNIPR}". Note that Binary \JUNIPR still learns
the  probabilities for type-$a$ and type-$b$ jets and still trains the same neural-network functions; however, it uses a more effective objective function for discrimination applications. We also note that training can easily be generalized to multiclass classification. 

As a test of the advantage that the binary objective function provides over its unary counterpart, we applied Binary \JUNIPR to the discrimination of quark- and gluon-jets. We used a mixed sample of $10^6$ \Pythia quark-jets and $10^6$ \Pythia gluon-jets from the data set at \texttt{energyflow.network} \cite{Komiske:2018cqr, komiske_patrick_2019_3164691}. We set aside $10^5$ jets of each type into a test set, $10^5$ for validation, and used the remaining 80\% of the jets for training.
For the \JUNIPR models, $P_\cJ(\text{jet}|\text{quark})$ and $P_\cJ(\text{jet}|\text{gluon})$, we used an LSTM of dimension 30 to model $h^{(t)}$ and separate feed-forward networks, each with a single hidden layer of dimension 10, to model $P_\text{end}$, $P_\text{mother}$, and $P_\text{branch}$.\footnote{The Binary \JUNIPR architecture is available at \url{github.com/andersjohanandreassen/JUNIPR} with example code. 
}

We began by pre-training the two \JUNIPR models using the  original unary objective function of \Eq{UnsuperObjective}. We followed the same training schedule as in \cite{JUNIPR}, but scaled down the number of epochs by a factor of 5 because this data set is larger than the one used there. Pre-training took about 5 hours on a 16-core CPU server for each model. 
After pre-training, we optimized the binary objective function of \Eq{SuperObjective} using Adam with standard settings \cite{kingma2014adam} and the following batch-size schedule: 
\begin{center}
 \begin{tabular}{c||cccccc} 
 Schedule ~&~ 1 epoch ~& 5 epochs ~& 10 epochs ~& 10 epochs  \\ 
 \hline
 batch size ~& 10 & 100 & 1000 & 2000
\end{tabular}
\end{center}
This segment of training took 12 hours on a 16-core CPU server. 
Binary \JUNIPR parameters were decided upon by evaluating the AUC (area under the ROC curve) on the validation set 10 times per epoch and choosing the model that achieved the maximal AUC during the final 10 training epochs. 
Note that different hyperparameters might be appropriate for different applications. 

\begin{figure}[t]
\centering
\includegraphics[width=\linewidth]{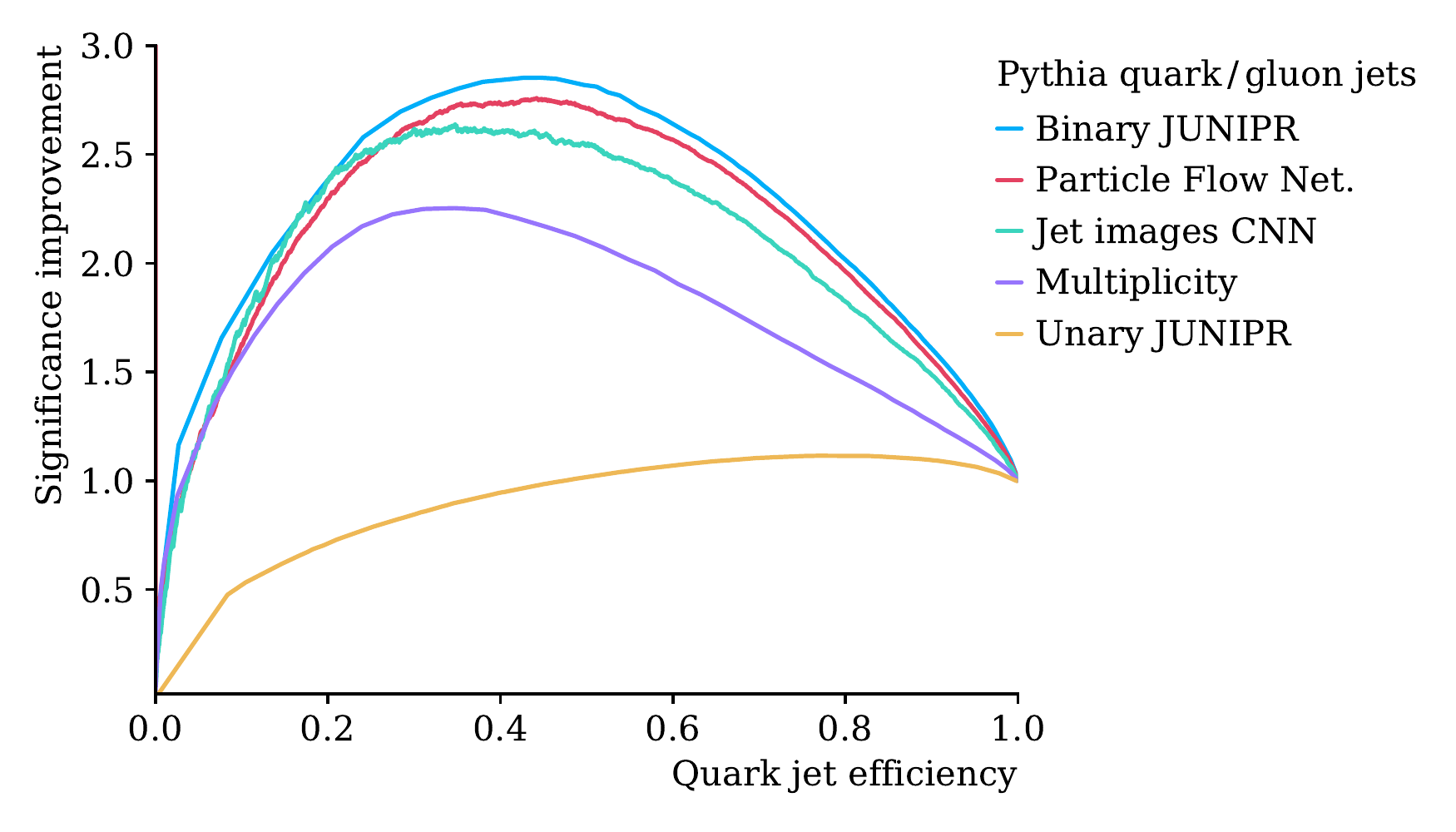}
\caption{Significance improvement $\frac{\varepsilon_Q}{\sqrt{\varepsilon_G}}$ as a function of $\varepsilon_Q$
for quark/gluon discrimination. Binary \JUNIPR is compared a Particle Flow Network \cite{Komiske:2018cqr}, a CNN using jet images \cite{Komiske:2018cqr}, constituent multiplicity, and unary \JUNIPR.}
\label{fig:QG_discrimination}
\end{figure}

In \Fig{QG_discrimination} we show the quark-versus-gluon Significance Improvement Curve~\cite{Gallicchio:2010dq}, (SIC), $\frac{\varepsilon_S}{\sqrt{\varepsilon_B}}$,  achieved by Binary \JUNIPR and compare it to recent results with previous
state-of-the-art discriminants: a CNN approach based on jet images~\cite{Komiske:2016rsd} (with architecture from \cite{Komiske:2018cqr}) and Particle Flow Networks \cite{Komiske:2018cqr}. One can see that Binary \JUNIPR offers a small-but-significant advantage.
Quantitatively, Binary \JUNIPR achieves an AUC of $0.8986 \pm 0.0004$, as compared to $0.8911 \pm 0.0008$ for Particle Flow Networks, and $0.8799 \pm 0.0008$ for the CNN. (Each reported number is the mean and semi-interquartile range over 10 trainings.) 
Unary \JUNIPR, trained with \Eq{UnsuperObjective}, performs significantly worse than the other methods, achieving an AUC of $0.6968 \pm 0.0008$. 
This demonstrates the importance of training \JUNIPR with the binary objective function of \Eq{SuperObjective} for classification.

As a second experiment, we trained and tested Binary \JUNIPR for boosted top-jet identification.
We used the same architecture and training schedule that were optimized for quark-versus-gluon discrimination. In doing so, we obtain a sense of the performance one might expect from Binary \JUNIPR without specialized hyperparameter tuning. The training, validation, and test data for this experiment are taken from \cite{Butter:2017cot}.
We found that untuned Binary \JUNIPR comes close to state-of-the-art top discrimination. Specifically, \JUNIPR achieves an AUC of $0.9810 \pm 0.0002$ as compared to $0.9819 \pm 0.0001$ attained using Particle Flow Networks~\cite{Komiske:2018cqr}, and 0.9848 reported for ParticleNet \cite{Qu:2019gqs}; all significantly outperform traditional boosted top-tagging methods~\cite{Kaplan:2008ie}. For a recent overview of machine learning in top tagging, see \cite{Kasieczka:2019dbj}.

Next we discuss the interpretability of \JUNIPR models.
As discussed below \Eq{IndividTimeStepModel}, each component of \JUNIPR's output has a well-defined physical meaning. 
Moreover, the output is structured along a physically-motivated binary tree, defined by clustering the momenta in a jet.
One can thus decompose \JUNIPR's prediction, say $P_\cJ(\text{top}\,|\,\text{jet})$ as in \Eq{binary_pred}, visually along the clustering tree.
In \Fig{jet_trees}, we show the clustering tree for an easily classifiable top jet drawn from the mixed top/QCD test set.
In the figure, we label the $t$-th node with $P^{(t)}(\text{top}\,|\,\text{jet})$, i.e.~the probability that the jet is top-type, given only the information present at branching step $t$;
this is computed with Binary \JUNIPR by substituting \Eq{IndividTimeStepModel} into \Eq{binary_pred}.
One can see, for example, that the 3-prong structure characteristic of $t \to W^+b \to u\,\bar d\,b$ contributes to large $P_\cJ(\text{top}\,|\,\text{jet})$. Quantitatively, this results in the two
hard branchings, with $P^{(t)}(\text{top}\,|\,\text{jet}) = 0.72$ and $0.71$, dominating the prediction.
By analyzing such trees, one can develop intuition for which branchings are most decisive in classifying different types of jets.

To be concrete, let us return to the Binary \JUNIPR model used to create \Fig{QG_discrimination}, which learned to discriminate quark and gluon jets from \Pythia. Much is already known about the difference between quark and gluon jets: gluon jets are known to be bigger, with larger multiplicity and larger shape parameters such as mass and width \cite{Gallicchio:2012ez,Gallicchio:2011xq}. Although many methods exist for quark/gluon discrimination, including other machine-learning approaches~\cite{Komiske:2016rsd,Fraser:2018ieu}, it is not clear how well these methods will work on actual data. In particular, it is known that real gluon jets are more similar to real quark jets than \Pythia leads us to believe~\cite{Aad:2014gea}. In particular, it is the modeling of gluon jets that seems most inaccurate. An alternative generator, \Herwig, produces and gluon jets that are more similar to its quark jets~\cite{Gras:2017jty}. Thus, we also considered a secondary challenge: determine how \Pythia and \Herwig gluon jets differ. To explore their differences, we trained a second Binary \JUNIPR model to discriminate \Pythia 8.226 and \Herwig 7.1.4 gluon jets using $10^6$ samples of each from \cite{komiske_patrick_2019_3164691, pathak_aditya_2019_3066475}.

\begin{figure*}
\centering
\includegraphics[width=0.32\linewidth]{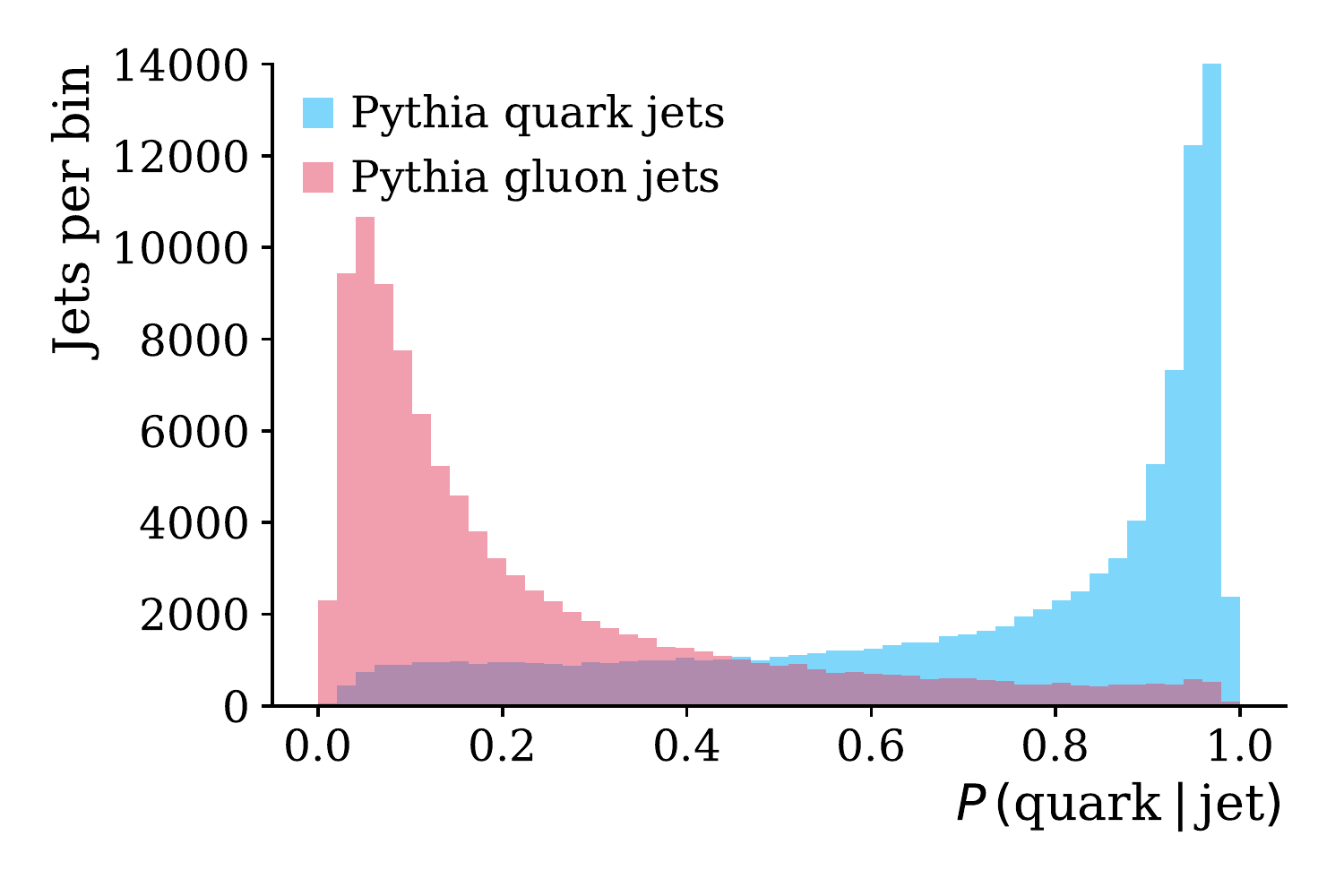}
\includegraphics[width=0.32\linewidth]{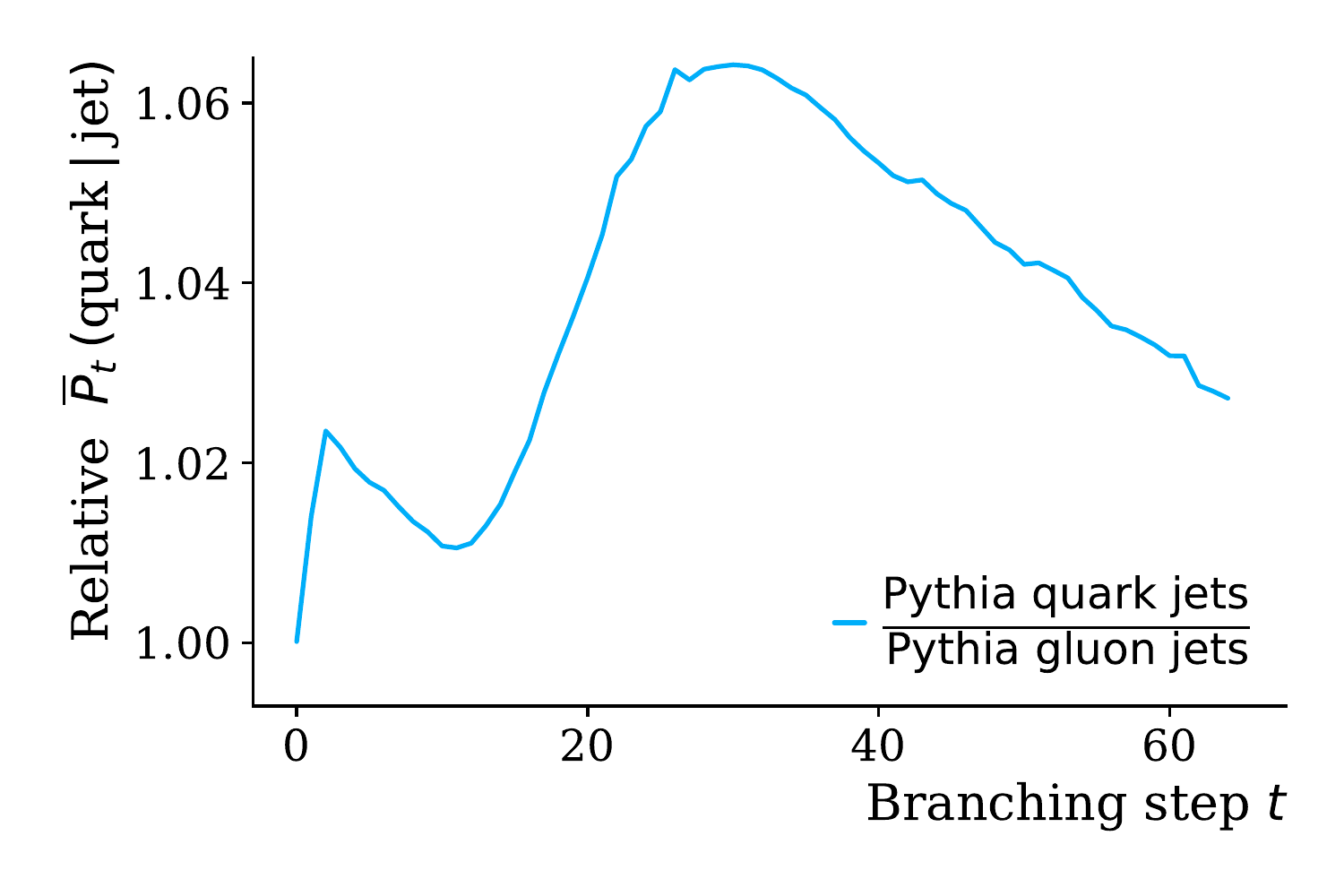}
\includegraphics[width=0.32\linewidth]{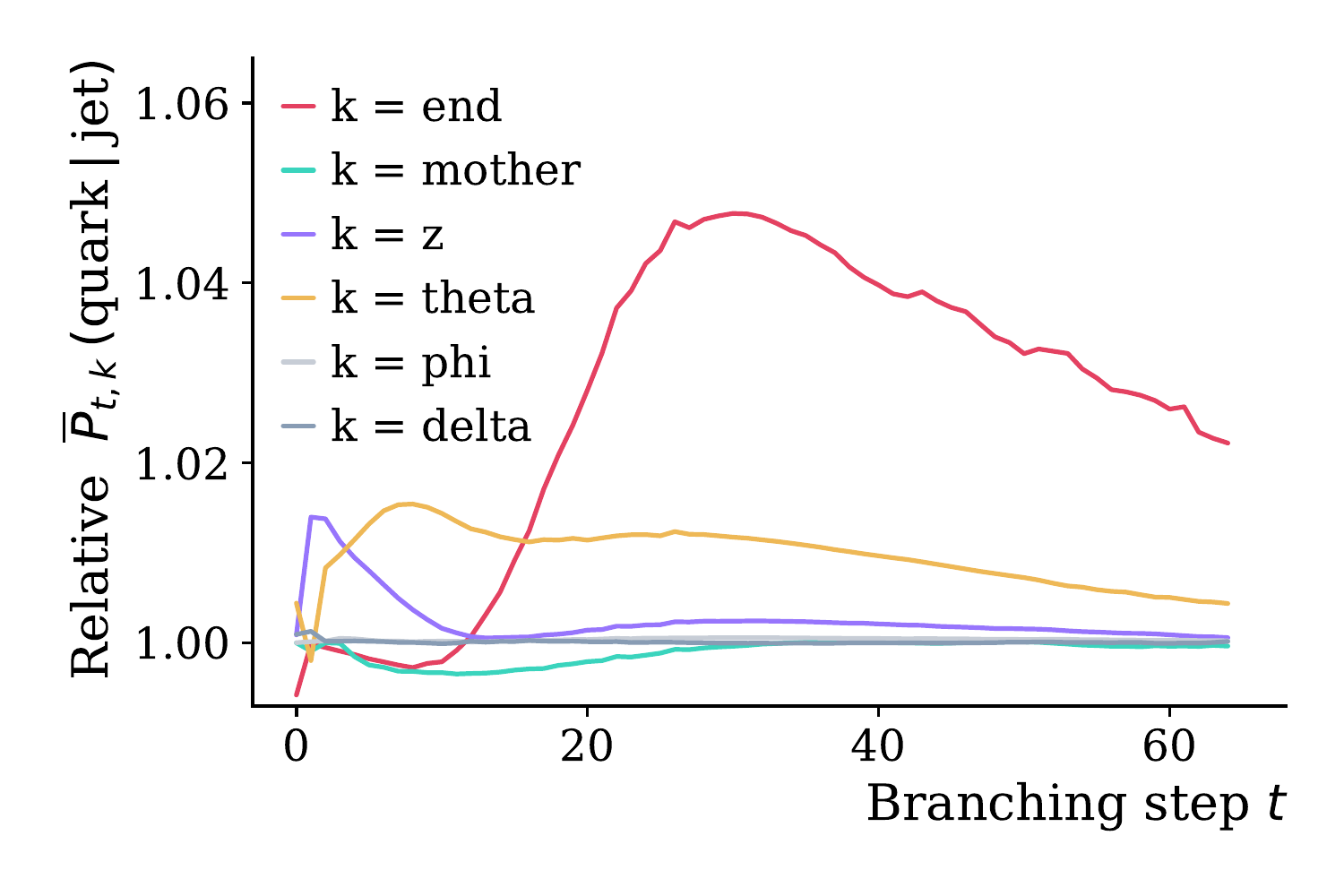}\\
\includegraphics[width=0.32\linewidth]{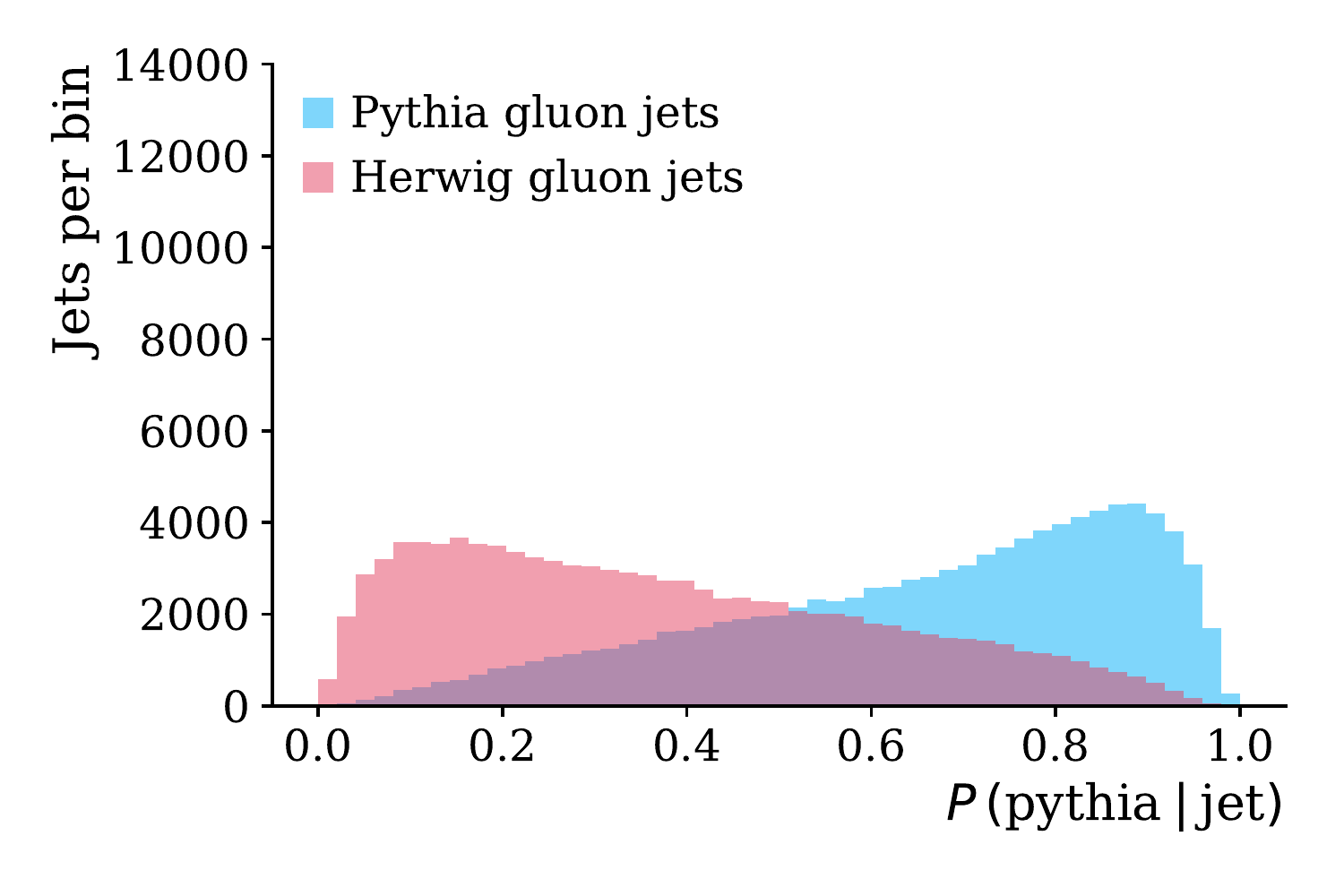}
\includegraphics[width=0.32\linewidth]{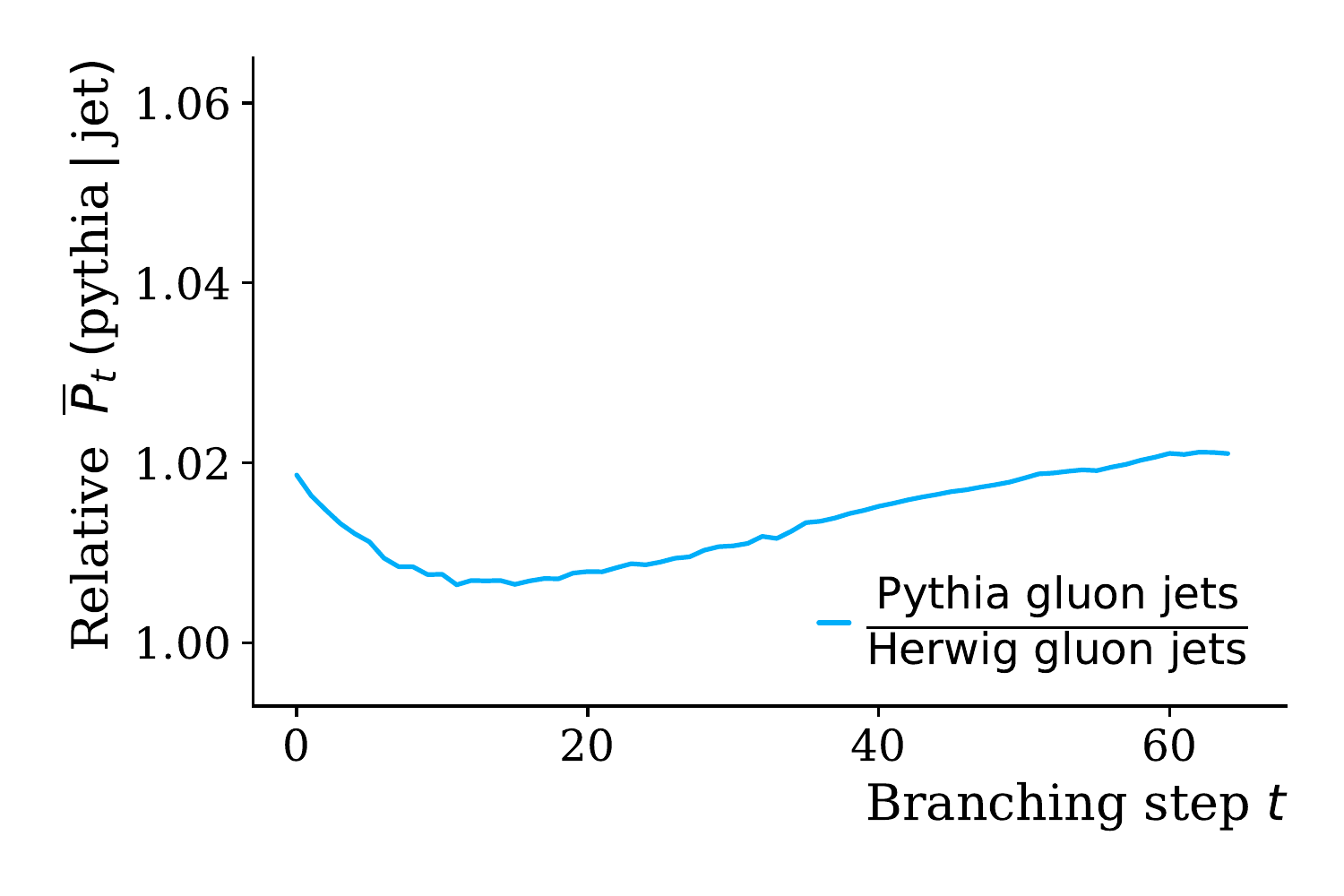}
\includegraphics[width=0.32\linewidth]{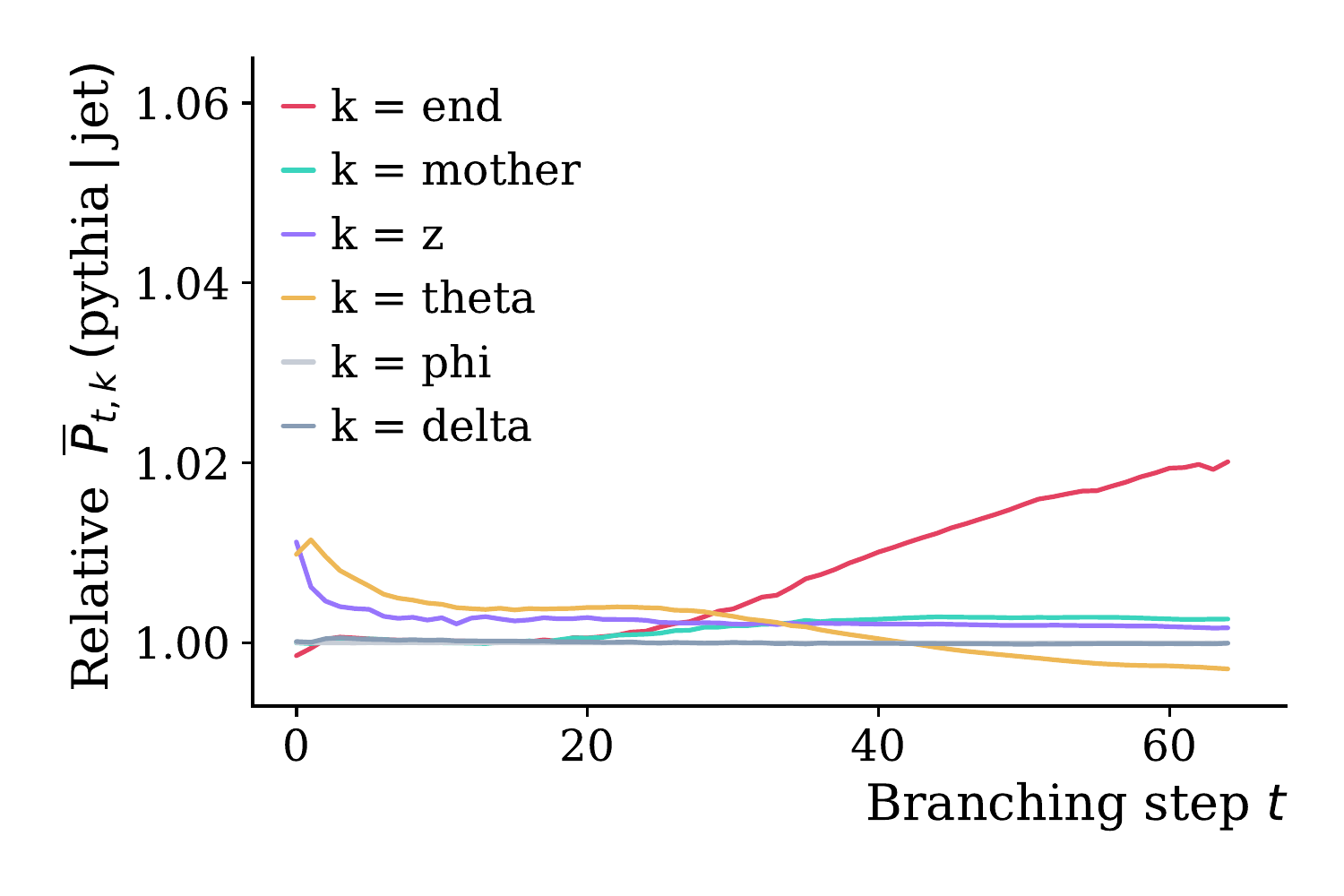}
\vspace{-2mm}
\caption{Quark/gluon (top row) and Pythia/Herwig discrimination (bottom row) with Binary \JUNIPR. Here we will refer to quark jets and \Pythia jets as ``signal'', and to gluon jets and \Herwig jets as ``background''. The left column shows the binary probability with which \JUNIPR predicts each jet is a signal jet. The middle column breaks these probabilities down by branching step in the clustering tree. Specifically, the plots show the ratio of $P_t(\text{signal}|\text{jet})$, averaged over signal jets in the numerator and background jets in the denominator. The right column breaks these ratios down further by branching component.}
\label{fig:interpretation}
\end{figure*}

\Fig{interpretation} shows another visualization, complementary to \Fig{jet_trees}, of exploring how \JUNIPR discriminates. 
The top row of \Fig{interpretation} shows how \JUNIPR  separates \Pythia quarks from \Pythia gluons, 
and the bottom row shows how \JUNIPR separates \Pythia gluons from \Herwig gluons.
In the middle column, the overall probability that \JUNIPR uses for discrimination is decomposed into branching steps $t$, averaged over all jets of the given class. 
From this, we see that near $t = 20$ -- 50 there is roughly 3 times the quark/gluon discrimination power per branching step as for $t = 1$ -- 10. 
This echoes the well-known fact that multiplicity allows one to separate quark and gluon jets better than perturbatively calculable observables sensitive to only the first few splittings \cite{Gallicchio:2012ez}. 
The lower-middle plot shows that differences in \Pythia and \Herwig gluon jets are more uniformly spread over branching steps. 

Not only can \JUNIPR break discrimination power down into branching steps; \JUNIPR can further decompose classification probability into components at each branching. These components are displayed in the right column of \Fig{interpretation}; there are discrete components, such as whether branchings should end, as well as the energy $z$ and angles $\theta, \phi, \delta$ of the branching itself. While multiplicity ($P_\text{end}$) is the main driver of performance for quark/gluon discrimination, the angle $\theta$ also contributes significantly over a wide range of branchings, echoing the importance of jet width in this context. For the \Pythia/\Herwig task, both the angle $\theta$ and energy fraction $z$ play a significant role in discrimination on early branchings, and multiplicity becomes important on later branchings. 

It is interesting that a significant fraction of the difference between \Pythia and \Herwig results from the way energy and angles are distributed early on in the clustering trees. 
Early branchings are controlled primarily by perturbative elements of the simulated parton showers. 
This suggests that a substantial portion of the difference between \Pythia and \Herwig gluon jets may be driven by the parton-shower implementations, rather than exclusively by the modelling of non-perturbative effects. 
To gain further insight into the importance of non-perturbative effects like hadronization in discrimination, \JUNIPR could be upgraded to include quantum numbers of final state particles --- a straightforward next step. 

In \cite{JUNIPR}, \JUNIPR was introduced as a new framework for unsupervised machine learning in particle physics that prioritizes interpretability. Given a jet, i.e.~a set of momenta, \JUNIPR learns to compute the probability of that jet, i.e.~how consistent the distribution of momenta is with the training data. In this paper, we used the same probabilistic framework as in~\cite{JUNIPR}, but we  augmented the training to learn subtle differences between two samples, an enhancement we call Binary \JUNIPR. We demonstrated both its effectiveness and interpretability, using quark/gluon jets, boosted top jets, and Monte-Carlo-generator-dependence as examples. It is satisfying that demanding interpretability does not lead to a loss in effectiveness: Binary \JUNIPR discriminates at levels competitive with the best machine-learning methods available. 

While these case studies were all simulation-based, there is a straightforward path to repeating these exercises on collider data. Although real data does not come with truth labels, there are established methods for working with mixed samples~\cite{Komiske:2018oaa,Metodiev:2017vrx} which can be adapted to \JUNIPR without much modification. Then one could use a data/simulation Binary \JUNIPR model to understand deficiencies in simulations.
One could also use insights derived from Binary \JUNIPR trees to judge whether predictions should be trusted experimentally (was information below experimental resolution deemed important?) or to design new calculable observables (sensitive to previously-overlooked decisive branchings). Having interpretable methods opens the door to whole new approaches to understanding data from particle colliders.

We thank P.~Komiske and A.~Parthak for assistance with the samples. This work is supported in part by the  U.S.~Department of Energy under contract DE-SC0013607.
Computations were performed on the CORI supercomputing resources at NERSC, Harvard's Odyssey cluster, and the Faculty Platform for machine learning.

\bibliography{JUNIPR_discrim.bib}
\bibliographystyle{utphys}

\end{document}